%% file: paper.tex
\documentclass[10pt,twocolumn]{article}

\usepackage[tmargin=1in,bmargin=1in,lmargin=0.75in,rmargin=0.75in,columnsep=0.33in]{geometry} 

\input{shared.tex}

\begin{document}

\date{}

\title{\Ttokens: a better coordination primitive\\for data-processing systems}

\author{
{\rm Andrea Lattuada}\\
ETH Zurich
\and
{\rm Frank McSherry}\\
Materialize Inc.
}

\maketitle

\begin{abstract}
Distributed data processing systems have advanced through models that expose more and more opportunities for concurrency within a computation. The \emph{scheduling} of these increasingly sophisticated models has become the bottleneck for improved throughput and reduced latency.

We present a new coordination primitive for dataflow systems, the \emph{\ttoken}, which minimizes the volume of information shared between the computation and host system, without surrendering precision about concurrency. Several projects have now used timestamp tokens, and were able to explore computational idioms that could not be expressed easily, if at all, in other platforms. Importantly, these projects did not need to design and implement whole systems to support their research.
\end{abstract}


\section{Introduction}

\begin{figure}[t]
\centering
\includegraphics[width=0.37\textwidth]{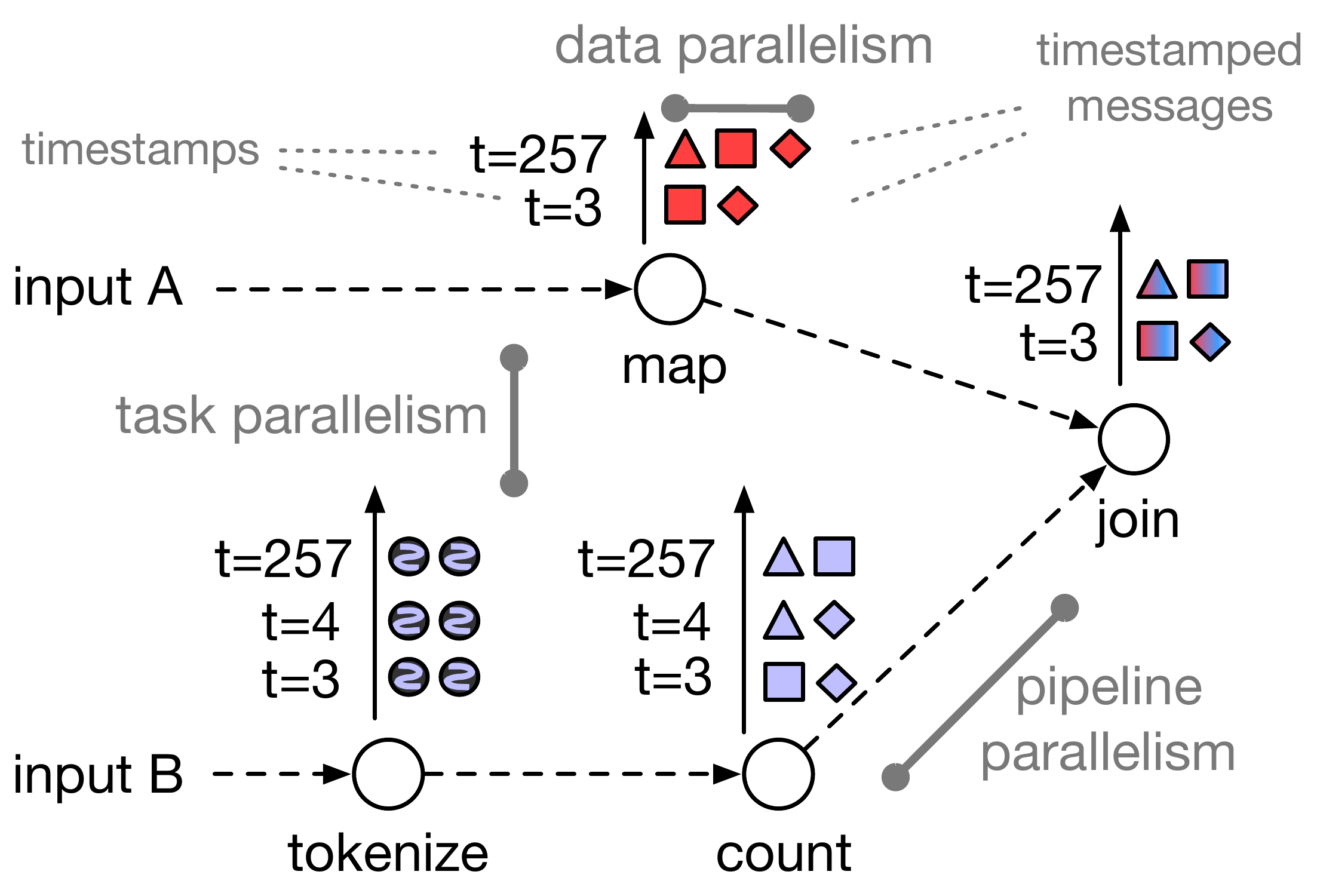}
\caption{ Data, task, and pipeline parallelism in dataflow systems with a temporal dimension. Data for each key is represented with different shapes. \texttt{map} transforms the tuples from input A: independent input tuples can be processed in a data-parallel fashion. \texttt{tokenize} acts on strings from input B, and \texttt{count} groups and tallies the tokens. \texttt{map} and \texttt{count} do not share a dataflow edge and are a candidate for task parallelism. \texttt{count} can process tokens at timestamp 4 while counts for timestamp 3 are \texttt{join}ed with the outputs of \texttt{map}: the temporal dimension enables pipeline parallelism between inter-dependent operators.}
\label{fig:parallelisms}
\end{figure}

Systems for data-intensive computation have advanced through programming models that allow programs to reveal progressively more opportunities for concurrency. Frameworks like MPI~\cite{site:mpi-forum} allow programmers only to explicitly sequence data-parallel computations. Systems like DryadLINQ~\cite{Yu:2008:DSGDDCUHL} and Spark~\cite{Zaharia:2012:RDDFAICC} use data-dependence graphs to allow programs to express task parallelism. Stream processors like Flink~\cite{Carbone:2015:AFSBPSE} and Naiad~\cite{Murray:2013:NTDS} (following \cite{site:AS, Akidau:2013:MFSPIS, Chandramouli:2009:OPDISQ}) add a temporal dataflow dimension to represent pipeline parallelism. In each case, new runtimes extract more detailed information about the computations, allowing them greater flexibility in their execution.  Figure~\ref{fig:parallelisms} demonstrates the forms of parallelism that can be expressed in these systems.

Dataflow systems have become limited by the complexity of the boundary between system and computation. Specifically, as computations provide progressively more fine-grained and detailed information about concurrency opportunities, the scalability and sophistication of the system schedulers must increase. In our experience, system complexity has increased to the point that scheduling rather than computation becomes the bottleneck that prevents higher throughputs and lower latencies.

System designers have the opportunity to reduce the \emph{volume} of coordination by reconsidering the interface between system and operator. For example, where Spark Streaming~\cite{Zaharia:2013:DSFSCS} must schedule distinct events to implement distinct logical times, Flink (and other stream processors) allow operators to retire batches of events corresponding to blocks of logical times, substantially improving throughput. Where Flink (and other stream processors) requires continual interaction with operators to confirm that they have no output at a logical time, Naiad asks operators to explicitly identify future times at which the operator should be notified, which is necessary to support cyclic dataflows. These interfaces reduce the volume of coordination, but require a deeper involvement of the system itself: continually invoking operators in Flink and sequencing notifications in Naiad.

We propose a simple dataflow coordination primitive, \textit{\ttokens}, which can dramatically simplify the design of advanced dataflow systems. Drawing inspiration from work on capability systems, a \ttoken is an in-memory object that can be held by an operator and provides the ability to produce timestamped data messages on a specific dataflow edge. A \ttoken does not require repeated interaction between system and operator to confirm, exercise, or release this ability. Instead, an operator accumulates and summarizes its interactions with its \ttokens. The system collects this information when most convenient, maintains a view of outstanding \ttokens, and provides summaries of potential input timestamps to each operator.  

\Ttokens make it relatively simple to introduce dataflow idioms that would be complicated or impossible in other systems. Although we have not previously reported on them since we designed and implemented them in an open-source data-processor, \ttokens have been in use for several years in various research and production projects.

Faucet~\cite{Lattuada:2016:FUMTFCDE} uses \ttokens to allow operators (and dataflow fragments) to implement their own flow control, without modifying system code.
DD \cite{McSherry:2020:SAPISSD} uses \ttokens to provide arbitrary granularity timestamps for differential dataflow, dramatically improving the throughput over the corresponding Naiad implementation.
Megaphone~\cite{Hoffmann:2019:MLSMDSD} uses \ttokens to specialize the implementations of operator-internal schedulers, for example using priority queues in operators that support them without requiring system-wide support.
In each case, \ttokens' separation between system and operators provided the flexibility to introduce behavior that would otherwise require the implementation of a specialized system.

\section{Coordination in dataflow systems}

A dataflow program is expressed as a directed graph $ (V, E) $ where nodes $ V $ represent data transformations and edges $ E $ represent the communication channels between the nodes. A dataflow system instantiates multiple workers and provides each with the dataflow graph. At runtime, the system exchanges data messages between workers, as the messages cross dataflow edges, and each worker independently applies data transformations in response to received data, producing output messages that are further exchanged and processed.

In modern dataflow systems, messages bear a logical timestamp $ t $, and dataflow operators maintain or advance timestamps as they process messages.
The system and operators collaborate to track outstanding messages by timestamp, so that operators can learn when certain input timestamps are complete and it is appropriate to produce the corresponding output.
Most commonly, the system provides the operator with a ``watermark'' or ``frontier'' indicating a lower bound on future timestamps the operator may observe, and the operator communicates to the system a lower bound on the timestamps it might still need to produce as output. The system is responsible for collecting and integrating the information from all operators, as well as the messages produced and retired, to provide correct lower bounds to the operators.

\subsection{Representative dataflow systems}

We now walk through several representative systems and relate their moving parts to dataflow coordination.

\smallskip
\textbf{Spark} models a computation as an acyclic dataflow graph, but without distinct logical times: inputs in Spark are either "complete" or "not yet complete". The Spark system tracks which inputs are complete and signals operators when their inputs are all complete and the operator can run to completion. Operators report back to the system as they complete their outputs.

\smallskip
\textbf{Flink} models computations as an acyclic dataflow graph, with integer logical times. Flink streams (dataflow edges) report an increasing integer ``watermark" lower-bounding the timestamps the stream may yet produce. These watermarks are interleaved in the stream of data itself, and each operator is required to produce them in their output streams as well. Flink does not have a centralized scheduler, and maintains a fresh view of its outputs only through the continued introduction of new watermarks in the dataflow inputs.

\smallskip
\textbf{Naiad} models computations as a potentially cyclic dataflow graph, with partially ordered logical times. Naiad operators request ``notifications" at specified logical times, and Naiad invokes a callback only once it determines that all messages bearing that logical time have been delivered. Naiad does not present operators with lower bounds for their inputs, and instead requires operators to defer the responsibility of scheduling to the system itself, in part because the logic for doing so requires a holistic view of the dataflow graph and all other pending notifications.

\smallskip

Each of these systems introduce new opportunities for concurrency, and corresponding performance gains on important tasks. However, no one system unifies the work of the others. We believe that unifying this work, and laying the groundwork for more advanced behaviors, requires a \emph{simplification} of the interface between system and operator, rather than further sophistication.


\section{\Ttokens}

We propose that dataflow systems and operator logic can coordinate precisely, efficiently, and ergonomically by explicitly handling in-memory tokens that represent their ability to produce outgoing data in the future. We borrow and adapt this idiom from capability systems (e.g. object-capability systems \cite{Dennis:1966:PSMC,Fabry:1974:CA}, capability-based protection and security \cite{Cohen:1975:PHOS,Mullender:1990:ADOS1}, hardware capabilities \cite{A.-Feustel:1973:ATA,Chisnall:2015:BPASMAM}). Similarly to capabilities\footnote{``Each capability [...] locates by means of a pointer some computing object, and indicates the actions that the computation may perform with respect to that object." \cite{Dennis:1966:PSMC}}, a \ttoken represents a computing object -- an operator output -- and the actions that can be performed with respect to that object: the production of data at timestamp $ t $ and dataflow location $ l $.

Following Naiad we refer to the pair of timestamp $ t $ and location $ l $ as a \textit{pointstamp} $ (t, l) $. A location can be either a node in $ V $ or an edge in $ E $.

\smallskip
\textbf{Definition.} 
A \textit{\ttoken} is a coordination primitive that names an associated pointstamp $ (t, l) $, and which gives its holder the ability to produce messages with timestamp $t$ at location $l$.

\smallskip \noindent
The location for a \ttoken is typically one of the output edges of the operator that holds it.

\smallskip

Nothwithstanding any other similarities to capabilities, our interest is in the information that holding \ttokens communicates to others. The system tracks the set of live \ttokens and summarizes this information to operators as \textit{frontiers}: lower bounds on the timestamps that operators may yet observe in their inputs. By downgrading (to future timestamps) or discarding their held \ttokens, operators allow frontiers to advance and the computation as a whole to make forward progress.

\subsection{The \ttoken life-cycle}

\begin{figure}[t]
\centering
\includegraphics[width=0.43\textwidth]{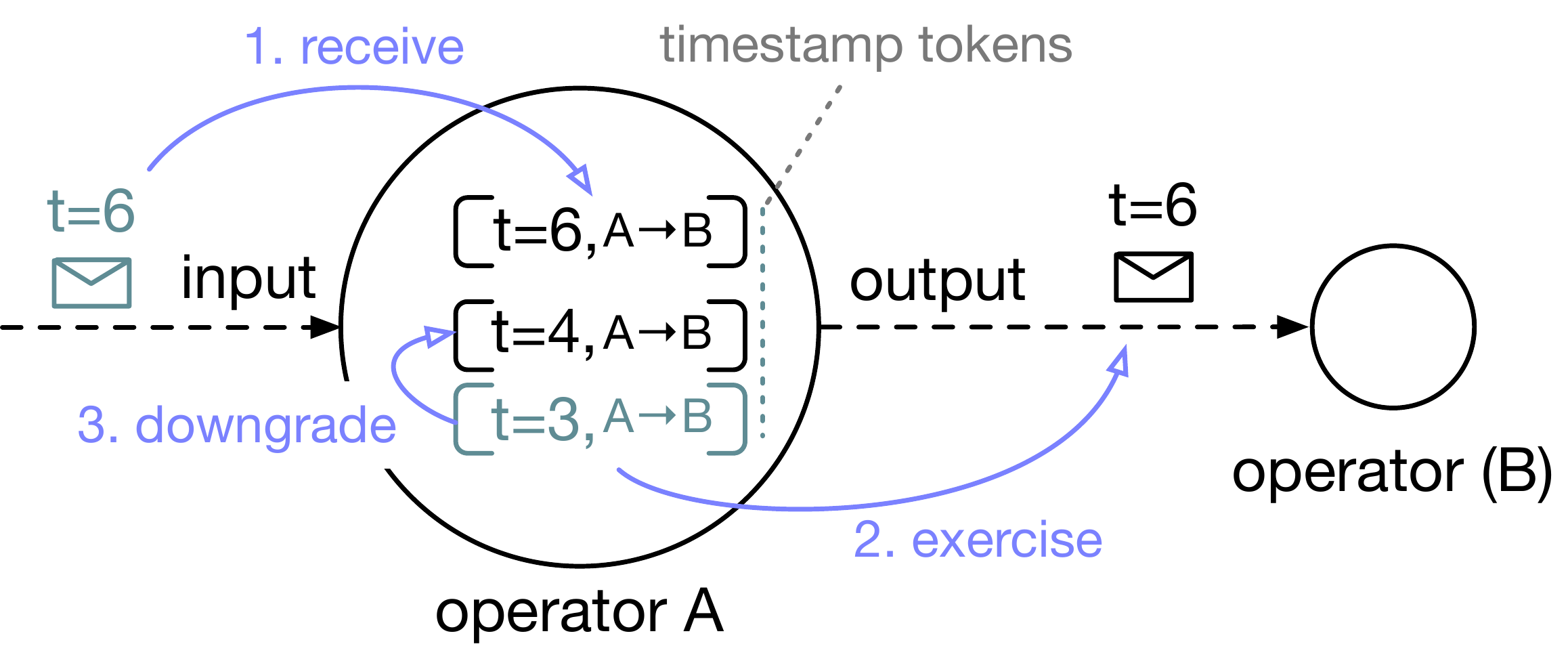}
\caption{\label{fig:lifecycle} \Ttoken life-cycle.}
\end{figure}

Each dataflow operator is initially provided with a \ttoken for each of its output edges, each bearing some minimal ``zero" timestamp. This gives each operator the opportunity to be a source of timestamped messages, even without receiving input messages. For many operators, their first actions will be to discard these \ttokens, by which they release their ability to produce output messages unprompted, and unblock the dataflow system at the same time. 

As a dataflow operator executes, it can receive, exercise, downgrade, and discard \ttokens (Figure~\ref{fig:lifecycle}). Operators receive timestamped input messages, each of which provides a \ttoken at that timestamp for each of the operator's outputs. Operators can produce timestamped output messages as long as they hold a \ttoken with the corresponding timestamp and output edge. Lastly, operators can arbitrarily hold, downgrade (to future timestamps), and discard their \ttokens as their logic dictates.

The dataflow system is informed of the net changes to the number of \ttokens for each pointstamp, but only passively in response to operator actions, rather than actively as a gatekeeper. Through this information the system can inform dataflow operators about the consequences of operator actions, without the specific details of the reasons for those actions.

\subsection{Coordination}

The coordination state of the dataflow system is the set of \ttokens, which when combined with the dataflow graph determines lower bounds for the timestamps at each operator input. As the set of \ttokens evolves these lower bounds advance, and the dataflow system has the responsibility of informing operators as this happens. 
The difference with \ttokens is that operators drive the \emph{production} of this information, instead of the system itself.\footnote{For example, Naiad does not allow operators to hold tokens across invocations; Timely Dataflow (without \ttokens) does, by allowing operators to participate directly (and often incorrectly) in the coordination protocol. Here, \ttokens are respectively more expressive, and safer.}

Operators have a great deal of flexibility in how (or even if) they respond to changes in their input frontiers (timestamp lower bounds). Certain streaming operators like \texttt{map} and \texttt{filter} can be oblivious to this information and process data as it arrives. Synchronous reduction operators like \texttt{reduce} should await the indication that they have received all inputs for a timestamp before they apply their reduction function and produce output. Hybrid operators like \texttt{count} may perform some accumulation in place and await the frontier advancing before producing the final tally for each timestamp. In each case the operator responds to input data and changes in its input frontiers, with output data and changes in its held \ttokens, but does not otherwise expose complexity to the system.


\section{Implementation}

We implemented \ttokens for Timely Dataflow~\cite{McSherry:TD} in the Rust programming language~\cite{site:RPL} \cite{Matsakis:2014:RL}. In our implementation, \ttokens are Rust types that wrap a timestamp, a location, and a bookkeeping data structure shared with the system. Operator logic manipulates \ttokens through their methods---cloning, downgrading, and dropping them---which update the shared data structure with integer \emph{changes} to the numbers of \ttokens at each timestamp and location. 

The timely dataflow system drains shared bookkeeping data structures outside of operator logic but on the same thread of control, which ensures the changes reflect atomic operator actions. Following Naiad's progress tracking protocol, these collected changes are broadcast among unsynchronized workers. Any subset of atomic updates forms a conservative view of the coordination state (the outstanding \ttokens) and is sufficient to maintain a conservative view of timestamp lower bounds for each operator across the otherwise asynchronous workers. 

The Rust~\cite{Weiss:2019:OER} language provides several features that simplify our implementation. Rust is type-safe, and users cannot fabricate \ttokens outside of unsafe code. Rust also does not allow users to destructure private \texttt{struct} fields, ensuring that we protect the shared bookkeeping data structure from direct user manipulation. Rust's affine type system ensures that users cannot casually copy \ttokens without explicit method calls, which allow us to interpose and increment counts. Finally, Rust eagerly invokes destructor logic, so that dropping a \ttoken is immediately visible to the system.

Our implementation is general enough to reproduce idioms from other systems, with no overhead. We have implemented Naiad notifications in library operator logic, and if in each invocation an operator processes only their least timestamp they reproduce Naiad's notification behavior. We can also implement Flink-style watermarks, with operators that explicitly hold \ttokens for their output watermarks and downgrade them whenever these watermarks advance.
Both these idioms are helpful but restrictive, and they are enforced system-wide in prior work.
Our intent is that operators should be able to choose the most appealing idiom, or new idioms as appropriate, without requiring the system to change as well.

This generality is not without some ergonomic cost: prior systems could more easily encourage operators make forward progress. Flink operators should eventually bring their output watermarks in line with their input watermarks, and Naiad operators should respond to notifications with something other than a re-notification request for the same time. From experience, user operators can more easily ``lose track'' of a \ttoken, for example when used as a key in a hash map and not discarded once its associated values have been processed. We use Rust's type system to raise the programmers awareness, by providing operators only a ``\ttoken option'', which the operator must then specifically retain to receive a \ttoken. Rust's lifetime system ensures at compile time that the options themselves can not be held by an operator, forcing it to explicitly retain or pass on \ttoken options.

\subsection{\Ttokens in code}

We present an extract of the main definitions of the \ttoken Rust API and implementation in \Cref{fig:listing-ttoken-api}.
A \textcode{TimestampToken}\clref{clnum:api-timestamptoken} wraps a timestamp\clref{clnum:api-time} and a bookkeeping data structure\clref{clnum:api-bookkeeping} shared with the system. These fields are private and the operator code cannot directly access or mutate them.
The bookkeeping data structure records the location for which the \textcode{TimestampToken} is valid, which will be checked by the system should the \textcode{TimestampToken} be exercised to send data. 
Operators may hold any number of \textcode{TimestampToken}s. 


Three methods, \textcode{downgrade}\clref{clnum:api-downgrade}, \textcode{clone}\clref{clnum:api-clone}, and \textcode{drop}\clref{clnum:api-drop}, are the only ways user code can directly manipulate the number of \ttokens at a pointstamp (without the use of Rust's \textcode{unsafe} keyword). The number of \ttokens at a pointstamp is indirectly manipulated by sending timestamped messages to the location of that pointstamp, through the \textcode{session}\clref{clnum:api-session} method.

Operator code can directly downgrade a \ttoken to a later timestamp with \textcode{downgrade}. This reduces the operator's ability to produce output at the wrapped timestamp, potentially to the point that the system can unblock downstream operators, though not beyond the timestamp downgraded \emph{to}. The implementation of \textcode{downgrade} updates the bookkeeping data-structure to inform the system of the net changes to the number of \ttokens for each pointstamp.

Operator code can also call into Rust's \textcode{clone} (deep copy) and \textcode{drop} (destructor) methods on \textcode{TimestampToken}. Custom implementations of these two methods respectively increment and decrement pointstamp counts for the wrapped timestamp in the bookkeeping data-structure. A \textcode{drop} call is automatically inserted by the Rust compiler whenever an object goes out of scope, and makes it much less likely that an operator will fail to release a \ttoken.

In order to transmit data along an output dataflow edge, an operator must \emph{express} a \ttoken. Access to outputs is guarded by an \textcode{OutputHandle}\clref{clnum:api-output-handle}, whose method \textcode{session}\clref{clnum:api-session} will create an active \textcode{Session} only when presented a reference to a \textcode{TimestampToken}. The \textcode{Session} is only valid for the wrapped timestamp, or timestamps greater than it. Rust's lifetime system ensures that the \textcode{TimestampToken} cannot be modified or dropped as long as the \textcode{Session} is active (until it is dropped). As long as the \textcode{Session} is available to user code, the \textcode{TimestampToken} is guaranteed to still exist unmodified. Sent data arrive at the destination bearing a \ttoken that can be used by the recipient.

\subsection{Ergonomic modifications}

The core \ttoken code is explained in the previous section, but we have also made several ergonomic improvements in at attempt to minimize the chance of unintended errors.

In addition to \textcode{TimestampToken} objects, which are \emph{owned} by the code and data structures that reference them, we also provide a \textcode{TimestampTokenRef} structure that cannot be held longer than a fairly narrow lexical scope. To acquire an owned token, user code must explicitly call \textcode{retain} which then results in a \textcode{TimestampToken}. We have found this reduces the incidences of user code unintentionally capturing and indefinitely holding a \ttoken, thereby stalling out dataflows.

Both \textcode{TimestampToken} and \textcode{TimestampTokenRef} implement a Rust trait \textcode{TimestampTokenTrait} that allows system code (specifically \textcode{session}) to accept either. This allows users to bypass the \textcode{retain} method and create a \textcode{Session} from a token reference, avoiding some syntax but importantly also avoiding bookkeeping when \ttoken ownership is not needed.

\Ttokens by default update shared bookkeeping data structures, but do not force the system to immediately act upon the changes they reflect. The operators that house an \textcode{OutputHandle} inform the system that it should consult the shared bookkeeping, when the operator yields control.
Several variants of \textcode{TimestampToken} take specific action when modified, including notifying the system that it should accept any updates and act on them. This allows these \ttokens to be used outside of the operators their pointstamps reference, and are especially useful for manual control of inputs to a dataflow when the logic cannot easily be encapsulated in an operator.

These modifications do not change the core behavior of \ttokens, but instead demonstrate how rough edges can be sanded down using layers atop \ttokens.

\begin{figure}[H]
  
\begin{lstlisting}[language=Rust]
/// The ability to send data with a
/// certain timestamp on a dataflow edge.
pub struct TimestampToken<T: Timestamp>#\clnum\label{clnum:api-timestamptoken}# {
    time#\clnum\label{clnum:api-time}#: T,
    bookkeeping#\clnum\label{clnum:api-bookkeeping}#: Bookkeeping<T>,
}

impl<T: Timestamp> TimestampToken<T> {
    /// The timestamp associated with this
    /// timestamp token.
    pub fn time#\clnum\label{clnum:api-time-fn}#(&self) -> &T { #\rm\dots# }

    /// Downgrades the timestamp token to
    /// one corresponding to `new_time`.
    pub fn downgrade#\clnum\label{clnum:api-downgrade}#(
        &mut self, new_time: &T) { #\rm\dots# }#\newline##\rm\dots#
}

impl<T: Timestamp> Clone#\clnum\label{clnum:api-clone}#
    for TimestampToken<T> {
    fn clone(&self) -> TimestampToken<T> { #\rm\dots# }
}

impl<T: Timestamp> Drop#\clnum\label{clnum:api-drop}##\newline#for TimestampToken<T> {
    fn drop(&mut self) { #\rm\dots# }
}#\newline##\rm\dots##\vspace{3pt}\hrule\vspace{3pt}#

impl<T: Timestamp, #\rm\dots#> OutputHandle<T, #\rm\dots#>#\clnum\label{clnum:api-output-handle}#
{
    /// Obtains a session that can send data
    /// at the timestamp associated with
    /// timestamp token `tok`.
    pub fn session#\clnum\label{clnum:api-session}#(
        &mut self, tok: &TimestampToken) -> Session<T, #\rm\dots#>#\newline#{ #\rm\dots# }
}
\end{lstlisting}
\caption{An extract of the \ttoken API and implementation in timely dataflow. We use circled letters, similar to \protect\circled{Z}, to mark points of interest in the code.}
\label{fig:listing-ttoken-api}

\end{figure}


\section{Example}
\label{sec:example}

\begin{figure}[th]
\centering
\includegraphics[width=0.47\textwidth]{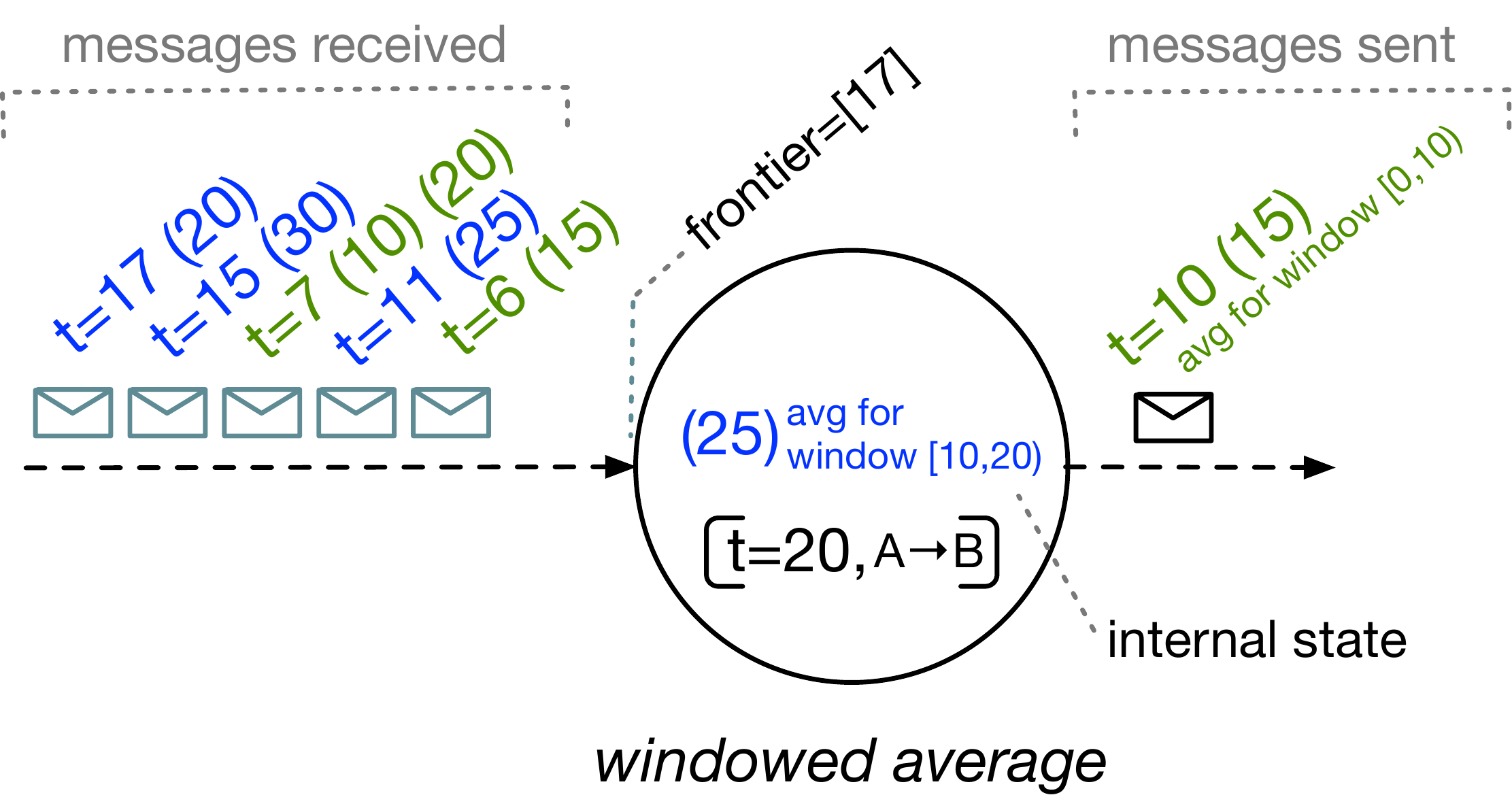}
\caption{The \textit{windowed average} operator in a sample execution with its internal state, a held \ttoken, and the data sent so far on its input and output edges.}
\label{fig:example-op-windowed-average}
\end{figure}

We use the example of \textit{tumbling windowed average} to demonstrate the life-cycle of \ttokens and how it generates coordination information. 
This operator receives timestamped integer-valued messages and reports the average every 10 timestamp units, at the timestamp of the start of the next window. The operator produces no output for windows which contain no data.
\Cref{fig:listing-tumbling-window} list the example code.

Importantly, this is code that one can write to introduce the behavior of a tumbling window to a system. It is not code that an end user should be expected to write each time they want a tumbling window. Rather, it can be written once, and then end users can simply invoke the method with appropriate parameters.


\Cref{fig:example-op-windowed-average} is a snapshot of the execution after the output for the window $ [0,10) $ has been produced. At this stage the operator maintains the current average for open windows (for which some data has been received but not necessarily all data) and a \ttoken to produce the output at the timestamp of the next open window (in the Figure, time \texttt{20}).

The operator has great flexibility in how it implements its specification. 
For example, the operator can choose to retain only the \ttokens for timestamps that are not greater than some other held \ttoken, reducing system interaction at the cost of local bookkeeping.
The operator can use ordered data structures to efficiently retire multiple windows at once, should the frontier advance suddenly.
The operator can maintain partial aggregations for out-of-order data while still being clear at which times they might emerge.

We walk through the sample code in \ref{sec:example-code} and call out the benefits \ttokens provide in \ref{sec:benefits}


\subsection{Example code}
\label{sec:example-code}

\begin{figure*}[t]
  
\begin{lstlisting}[
    language=Rust,
    numbers=left,
    numbersep=8pt,                   % how far the line-numbers are from the code
    numberstyle=\color{gray},
    xleftmargin=2em,
    framexleftmargin=2em
]
/// User-defined structure to maintain window data.
struct WindowData#\clnum\label{clnum:tumbling-struct-windowdata}# { pub sum: u64, pub count: u64 }
pub fn singleton_frontier(frontier: &MutableAntichain<u64>) -> u64 {
    frontier.frontier().first().cloned().unwrap_or(u64::MAX)
}#\newline##\rm\dots#
// The `unary_frontier` method defines a new operator from a anonymous function that specifies its logic.
stream.unary_frontier(
  Exchange::new(|x| x % (peers as u64)), "tumbling window", #\clnum\label{clnum:tumbling-outer-anonymous-function}#|#\clnum\label{clnum:tumbling-init-tok}#tok, _info| {
  #\clnum\label{clnum:tumbling-init-tok-assert}#assert!(*tok.time() == 0);
  #\clnum\label{clnum:tumbling-init-drop}#std::mem::drop(tok);
  let mut windows: BTreeMap<u64, (TimestampToken<u64>, WindowData)>#\clnum\label{clnum:tumbling-init-data}# = BTreeMap::new();
  // Define the anonymous function that is repeatedly invoked with input and output handles.
  #\clnum\label{clnum:tumbling-inner-anonymous-function}#move |input, output| {
      for (tok_ref#\clnum\label{clnum:tumbling-tok-ref}#, batch) in input {#\clnum\label{clnum:tumbling-input-for-each}#
          #\clnum\label{clnum:tumbling-window-ts}#let window_ts = round_up_to_multiple(*tok_ref.time(), WINDOW_SIZE);
          if !windows.contains_key(&window_ts) {#\clnum\label{clnum:tumbling-new-window-if}#
              let mut window_tok = #\clnum\label{clnum:tumbling-captured-tok}#tok_ref.retain();
              window_tok.downgrade(&window_ts);
              windows.insert(window_ts, (window_tok, WindowData { sum: 0, count: 0 }));
          }
          let (_, ref mut window_data#\clnum\label{clnum:tumbling-window-ref}#) = windows.get_mut(&window_ts).unwrap();
          for d in batch {
              window_data.sum += *d; window_data.count += 1;
          }
      }
      let target_ts = singleton_frontier(input.frontier()#\clnum\label{clnum:tumbling-frontier}#);
      for (_, (tok#\clnum\label{clnum:tumbling-range-tok}#, window)) in windows.range(0..target_ts) {#\clnum\label{clnum:tumbling-range}#
          #\clnum\label{clnum:tumbling-session-give}#output.session(&tok#\clnum\label{clnum:tumbling-session-tok-ref}#).give(window.sum as f64 / window.count as f64);
      }
      #\clnum\label{clnum:tumbling-windows-remove}#windows.remove_range(0..target_ts);
  }
})
\end{lstlisting}
\caption{A possible implementation of the tumbling window average operator described in \Cref{sec:example}. We use circled letters, similar to \protect\circled{Z}, to mark points of interest in the code.}
\label{fig:listing-tumbling-window}

\end{figure*}

\Cref{fig:listing-tumbling-window} shows the code listing for one of the many possible implementation of the \textit{tumbling windowed average} operator described in \Cref{sec:example}. The code presented closely resembles the real implementation of the operator, with some minor syntax modifications to aid readability and avoid Rust-isms that can be unfamiliar to the reader. Although detailed, this is the implementation expected of the system implementor; we expect end users would then access this functionality through a layer of abstraction rather than write it themselves.

The outer anonymous function\clref{clnum:tumbling-outer-anonymous-function} is invoked once by the system to intitialize the operator with a default \ttoken\clref{clnum:tumbling-init-tok} at time \texttt{0}\clref{clnum:tumbling-init-tok-assert}, which is immediately dropped\clref{clnum:tumbling-init-drop}. The operator initializes an ordered map\clref{clnum:tumbling-init-data} to store partial state for open windows: the timestamp for the end of the window maps to a tuple carrying the corresponding \ttoken and the partial \textcode{WindowData}\clref{clnum:tumbling-struct-windowdata} (the partial sum and count).

The inner anonymous function\clref{clnum:tumbling-inner-anonymous-function} contains the operator logic that is invoked every time the operator is scheduled. For each batch of input messages at a certain timestamp\clref{clnum:tumbling-input-for-each}, it computes the end-of-window timestamp\clref{clnum:tumbling-window-ts} from the message timestamp wrapped in the \ttoken{ }\textcode{tok\_ref}\clref{clnum:tumbling-tok-ref} (in the form of a \textcode{TimestampTokenRef}). If it has not seen data for this window before\clref{clnum:tumbling-new-window-if}, it captures\clref{clnum:tumbling-captured-tok} the \ttoken, immediately downgrades it to the end-of-window timestamp, and stores it along with initialized empty window data into the \textcode{windows} map.

The \ttokens stored in the map implicitly inform the coordination state of the operator: the system is informed of pointstamp changes after each invocation of the operator logic caused by \textcode{retain}, \textcode{downgrade}, and \textcode{drop} (when a \ttoken is finally removed from the map and dropped).

For each batch of input messages the operator logic obtains a mutable reference\clref{clnum:tumbling-window-ref} to the corresponding window data in the map, and updates the partial sum and count with each data point. Processing of new input concludes here.

The operator logic then needs to determine which windows have closed and emit the computed averages for them. This information is based on the set of live \ttokens in the system and is summarized by the system as per-input \textit{frontiers} at each operator: \textcode{input.frontier()}\clref{clnum:tumbling-frontier}. In general, timestamps in timely dataflow can be multidimensional and result in frontiers defined by multiple minima, but in this case we know that timestamps, and consequently frontiers, are represented by a single unsigned integer value. The frontier value represents the lower bound on timestamps that may still appear on the input: consequently we can safely retire all windows with end-of-window timestamps up to, but excluding, the frontier (\textcode{target\_ns}).

We leverage the map order to iterate over all open windows up to \textcode{target\_ns}\clref{clnum:tumbling-range}, and because we stored \ttokens alongside the window data, we obtain them during iteration\clref{clnum:tumbling-range-tok} and can immidiately leverage them to emit the computed averages at the correct timestamps\clref{clnum:tumbling-session-give}: to do so, we are required to pass in a reference to the \ttoken\clref{clnum:tumbling-session-tok-ref}. This ensures at compile time that the operator logic has the capability to send data at a certain timestamp.

The operator logic finally drops from the map all the windows it has just processed\clref{clnum:tumbling-windows-remove}. The \textcode{drop} code for the \ttokens stored in the values removed from the map are invoked automatically (and eagerly): this again updates the pointstamp changes that are reported to the system, and ensures that the frontiers for other downstream operators are updated accordingly.

\subsection{Benefits}
\label{sec:benefits}

The operator implementation above has several benefits that are prevented in other systems.

In a Spark-like system, where an operator is scheduled for each distinct timestamp, the operator would be unable to retire blocks of times concurrently. 
This limitation harms the throughput of data loading, and lowers the operator's throughput when bursts of differently timestamped data arrive. With \ttokens entire intervals of time can be closed at once, and the operator can perform all consequent work concurrently.

In a Flink-like system, the operator must be continually interrogated to advance its output watermark. Even if the operator input is idle for periods of time, the operator must remain active to inform downstream operators that there is no data. This scenario is more common than it might seem, with monitoring applications like fraud detection in which one wants to quickly confirm the absence of results. With \ttokens the system can bypass the operator entirely, reducing compute load and the critical path latency.

In a Naiad-like system, the operator must defer scheduling to the system. Should a batch of times be retired at once, as when a watermark finally arrives, the operator must repeatedly yield to the system and be re-invoked with advancing timestamps. With \ttokens the operator can perform this work on its own, using an efficient ordered data structure.

In addition, \ttokens avoid restrictions on dataflow structure, for example the requirement (seen in Spark and Flink) that dataflow graphs be acyclic.
Each of these benefits derive from involving the system \emph{less}, instead providing the operator with both more information and more agency. 


\section{Building with \ttokens}
\label{sec:stories}

\Ttokens have been in use for several years. In this section, we relate examples where developers and researchers found \ttokens to be especially helpful in
building frameworks that implement new dataflow programming patterns. In each case, \ttokens and specialized operator logic allowed projects to avoid re-implementing parts of the timely dataflow system itself.

\subsection{Co-operative control flow}

Dataflow operators may run for a long time or produce large amounts of output data, and should yield control so that other operators can execute and potentially retire some of the output data.
However, Naiad's execution model asks an operator to run to completion for each notification, and the return of control is an indication that the operator has completed its task.
\Ttokens allow operators to yield control without yielding the right to resume execution and produce output in the future.

Faucet~\cite{Lattuada:2016:FUMTFCDE} uses \ttokens to implement \emph{user-level flow control}. 
This mechanism supports dataflow operators that may produce unboundedly large numbers of output messages for each input. Faucet operators produce outputs up to a certain limit and then yield control until these messages are retired. 
Whenever an operator yields due to a reached limit, it retains the \ttoken to indicate it has further output to produce.
This design allowed the Faucet authors to implement flow control in user code, without requiring modifications to the underlying system.

\subsection{Fine-grained timestamps}

Systems that track real time may process events with timestamps denominated in nanoseconds. Naiad assumes responsibility for ordering all events with distinct timestamps, and for high-resolution timestamps this can overwhelm the system.
\Ttokens provide a mechanism for the operator to determine the granularity at which it reports outstanding timestamps to the system, without involving the system in each timestamp that is processed.

In DD~\cite{McSherry:2020:SAPISSD}, each event has a potentially unique timestamp, and operators receive and must react to a stream of such events. 
Rather than present each timestamp to the timely dataflow system, DD's operator implementations batch messages into ``intervals''.
An operator retains the least \ttokens for the times of un-batched messages it holds, and as the operator's frontier advances the operator creates new batches containing all events whose timestamps are not in advance of the new frontier.
The operator uses its current \ttokens to produce any output corresponding to the batch, and then downgrades its \ttokens once, to the new lower envelope of its un-batched messages.
This design allows the operators to interact with the host timely dataflow system at a coarse granularity, independent of the timestamp granularity.

\subsection{Optimized scheduling}
\label{sec:stories-scheduling}

Timely dataflow computations may act on general partially ordered timestamps, and with large numbers of outstanding events it may be unclear which events should be processed next.
A system like Naiad stores all events in an unsorted list and performs a sequential pass through this list in each scheduling round, limiting the minimum latency.
Alternately, stream processors that only act on totally ordered timestamps can use priority queues to quickly extract only the relevant events.
\Ttokens provide operators the ability to organize their schedulable work themselves, without pushing their implementation into the system itself.

In Megaphone~\cite{Hoffmann:2019:MLSMDSD}, a migration mechanism for timely dataflow, its authors implemented the NEXMark benchmark which contains a variety of streaming computations, and in particular a variety of \emph{windowed} computations.
These computations have timestamps that are denominated in nanoseconds, and in one case a windowed computation with a 12 hour continuous slide (and so, an effectively unbounded number of distinct timestamps in play at any time).
Their implementation uses priority queues of \ttokens to schedule the work in these specific operators, providing millisecond latencies without compromising the ability of the rest of the system to handle partially-ordered timestamps.

\subsection{Formalisation and safety proof}

\Ttokens express a clean interface between operators and the system.
An effort to formalize and verify the safety of the core coordination component of timely dataflow\cite{DBLP:conf/itp/0002DLT21} uses \ttokens as a basis for the formalization: the authors can precisely model what actions each instance of an operator can perform, in contrast with previous formalisation work\cite{DBLP:conf/forte/AbadiMMR13} that pre-dates \ttokens.


\section{Evaluation}
\label{sec:eval}

Our hypothesis is that by reducing systems complexity and granting more control on scheduling to individual operators, \ttokens remove the scheduling bottleneck that prevents modern data processing systems from reaching higher throughputs and lower latencies. We evaluate this hypothesis with a set of microbenchmarks designed to compare the different coordination mechanisms in prior art with \ttokens (\Cref{sec:eval-micro} and \Cref{sec:eval-opsequence}) and with more complex workloads that attempt to replicate real-world operating conditions (\Cref{sec:eval-nexmark}).
We hope to observe that \ttokens operate robustly in all settings where any coordination mechanism avoids collapse, and is never substantially worse than the best coordination mechanism. 

We compare \ttokens against the Naiad-style notification API already available in Timely Dataflow. In order to compare with Flink-style watermarks without the confounding factor of running on a different platform (like Flink's), we re-implemented Flink's watermarks technique on the same communication and scheduling framework provided by Timely Dataflow. In some of the experiments (\Cref{fig:wordcount-quantum,fig:wordcount-scaling}), where the technique selected has limited impact on performance, \ttokens and Flink-style watermarks achieve nearly identical latency, showing that our implementation does not unfairly disadvantage watermarks.

We observe that \ttokens avoid the collapse that notifications experience for high numbers of distinct timestamps (\Cref{fig:wordcount-scaling}), and the collapse that watermarks experience for complex dataflows (\Cref{{fig:opsequence}}). In all cases, \ttokens remain among the best approaches.

\subsection{Experimental setup}

We run all experiments on a CloudLab\cite{CloudLab} server with one AMD EPYC 7452 with 32 physical cores and 128GB of RAM. 
We disable simultaneous multi-threading (SMT) and 
we pin each timely dataflow worker to a distinct physical core.

Our open-loop testing harness supplies the input at a specified rate, even if the system itself becomes less responsive. We record the observed latency in units of nanoseconds in a histogram of logarithmically-sized bins. If the system becomes overloaded and end-to-end latency becomes greater than 1 second, the testing harness regards the experiment as failed.


\subsection{Microbenchmarks}
\label{sec:eval-micro}

Our microbenchmarks use a simple dataflow program that consists of a single stateful operator that computes the overall rolling count of unique words observed on the inputs. Every time the operator receives a word, it updates the internal count, and sends an output message with the updated value.

To determine the effectiveness of handling fine-grained timestamps with various techniques, we generate input at a given constant rate and assign different timestamps to each input tuple based on when it was generated. The assigned timestamps are quantized to powers-of-two ranging from $2^{8}$ to $2^{16}$ nanoseconds (``ns'' in the following). A timestamp quantum of $2^{x}$ns means that regardless of the input rate, there can be at most
$ \frac{1\times10^{9}}{2^{x}} $
distinct timestamps in the ingested data per second. For example, with a timestamp quantum of $2^{8}$ns ($256$ns), at most ~4 million timestamps per second can be generated.

Varying the size of the quantum allows us to evaluate how well a mechanism can handle coarser or finer timestamp granularities. With a smaller timestamp quantum, the system can provide higher time resolution in the output it produces. As previously discussed, with Naiad-style notifications, the operator needs to interact with the system for each logical time it processes, and for which it requires a notification.

\newcommand{\expwordcountquantumwatermarks} {experiments/wordcount-quantum/epyc7452-c9e6f1d-w8-quantums-watermarks.txt}
\newcommand{\expwordcountquantumnotificator}{experiments/wordcount-quantum/epyc7452-2ac8dfc-2-w8-quantums-notificator.txt}
\newcommand{\expwordcountquantumtokens}     {experiments/wordcount-quantum/epyc7452-2ac8dfc-2-w8-quantums-tokens.txt}

\newcommand{\expwordcountrateperworkertwo}{8000000}

\begin{figure}[tbh] 
  \centering
  \begin{tikzpicture}
    \begin{groupplot}[
      group style={
        group size=1 by 3,
        horizontal sep=0.2cm,
        vertical sep=0.6cm,
        xticklabels at=edge bottom,
        yticklabels at=edge left
      },
      xmin=7,
      xmax=21,
      width=.5\linewidth,
      height=.44\linewidth,
      x tick label style={/pgf/number format/.cd, fixed, precision=0},
      log number format basis/.code 2 args={\pgfmathparse{#1^(#2)}\pgfmathprintnumber{\pgfmathresult}ms},
      max space between ticks=20,
      legend style={at={(0,1)},anchor=south west,font=\small,draw=none},
      ]

      \nextgroupplot[
        ymode=log,
        ymin=0.5,
        ymax=10,
        ylabel=max latency
      ]
      \draw[ultra thick, draw=blue] (axis cs: 8,5) -- node[above,text=blue]{\footnotesize DNF} (axis cs: 12,5);
      \addplot+[dashdotted,olive,mark=diamond,mark options={solid,fill=olive},discard if not={rate_per_worker}{4000000}] table [x expr=log2(\thisrow{quantum}),y expr=\thisrow{max} / 1000000] {\expwordcountquantumwatermarks};
      \addplot+[dashed,blue,mark=o,mark options={solid,fill=blue},discard if not={rate_per_worker}{4000000}] table [x expr=log2(\thisrow{quantum}),y expr=\thisrow{max} / 1000000] {\expwordcountquantumnotificator};
      \addplot+[solid,red,mark=square,mark options={solid,fill=red},discard if not={rate_per_worker}{4000000}] table [x expr=log2(\thisrow{quantum}),y expr=\thisrow{max} / 1000000] {\expwordcountquantumtokens};

      \legend{watermarks, notifications, tokens}




      \nextgroupplot[
        ymode=log,
        ymin=0.1,
        ymax=8,
        ylabel=p999 latency
      ]
      \draw[ultra thick, draw=blue] (axis cs: 8,1) -- node[above,text=blue]{\footnotesize DNF} (axis cs: 12,1);
      \addplot+[dashdotted,olive,mark=diamond,mark options={solid,fill=olive},discard if not={rate_per_worker}{4000000}] table [x expr=log2(\thisrow{quantum}),y expr=\thisrow{0.999} / 1000000] {\expwordcountquantumwatermarks};
      \addplot+[dashed,blue,mark=o,mark options={solid,fill=blue},discard if not={rate_per_worker}{4000000}] table [x expr=log2(\thisrow{quantum}),y expr=\thisrow{0.999} / 1000000] {\expwordcountquantumnotificator};
      \addplot+[solid,red,mark=square,mark options={solid,fill=red},discard if not={rate_per_worker}{4000000}] table [x expr=log2(\thisrow{quantum}),y expr=\thisrow{0.999} / 1000000] {\expwordcountquantumtokens};

      \nextgroupplot[
        ymode=log,
        xlabel style={align=center, text width=.30\linewidth},
        xlabel={quantum ($2^x$ns)\\32M tuples/sec},
        ymin=0.1,
        ymax=2,
        ylabel=median latency
      ]
      \draw[ultra thick, draw=blue] (axis cs: 8,1) -- node[above,text=blue]{\footnotesize DNF} (axis cs: 12,1);
      \addplot+[dashdotted,olive,mark=diamond,mark options={solid,fill=olive},discard if not={rate_per_worker}{4000000}] table [x expr=log2(\thisrow{quantum}),y expr=\thisrow{0.5} / 1000000] {\expwordcountquantumwatermarks};
      \addplot+[dashed,blue,mark=o,mark options={solid,fill=blue},discard if not={rate_per_worker}{4000000}] table [x expr=log2(\thisrow{quantum}),y expr=\thisrow{0.5} / 1000000] {\expwordcountquantumnotificator};
      \addplot+[solid,red,mark=square,mark options={solid,fill=red},discard if not={rate_per_worker}{4000000}] table [x expr=log2(\thisrow{quantum}),y expr=\thisrow{0.5} / 1000000] {\expwordcountquantumtokens};

    \end{groupplot}
  \end{tikzpicture}
  \begin{tikzpicture}
    \begin{groupplot}[
      group style={
        group size=1 by 3,
        horizontal sep=0.2cm,
        vertical sep=0.6cm,
        xticklabels at=edge bottom,
        yticklabels at=edge left
      },
      xmin=7,
      xmax=21,
      width=.5\linewidth,
      height=.44\linewidth,
      x tick label style={/pgf/number format/.cd, fixed, precision=0},
      log number format basis/.code 2 args={\pgfmathparse{#1^(#2)}\pgfmathprintnumber{\pgfmathresult}ms},
      max space between ticks=20,
      legend columns=-1,
      legend style={at={(0,1)},anchor=south west,font=\small,draw=none},
      ]
      \nextgroupplot[
        ymode=log,
        ymin=0.5,
        ymax=10
      ]
      \draw[ultra thick, draw=olive] (axis cs: 8,2) -- node[above,text=olive]{\footnotesize DNF} (axis cs: 10,2);
      \draw[ultra thick, draw=blue] (axis cs: 8,.8) -- node[above,text=blue]{\footnotesize DNF} (axis cs: 13,.8);
      \draw[ultra thick, draw=red] (axis cs: 8,5) -- node[above,text=red]{\footnotesize DNF} (axis cs: 11,5);
      \addplot+[dashdotted,olive,mark=diamond,mark options={solid,fill=olive},discard if not={rate_per_worker}{\expwordcountrateperworkertwo}] table [x expr=log2(\thisrow{quantum}),y expr=\thisrow{max} / 1000000] {\expwordcountquantumwatermarks};
      \addplot+[dashed,blue,mark=o,mark options={solid,fill=blue},discard if not={rate_per_worker}{\expwordcountrateperworkertwo}] table [x expr=log2(\thisrow{quantum}),y expr=\thisrow{max} / 1000000] {\expwordcountquantumnotificator};
      \addplot+[solid,red,mark=square,mark options={solid,fill=red},discard if not={rate_per_worker}{\expwordcountrateperworkertwo}] table [x expr=log2(\thisrow{quantum}),y expr=\thisrow{max} / 1000000] {\expwordcountquantumtokens};

      \nextgroupplot[
        ymode=log,
        ymin=0.1,
        ymax=8,
      ]
      \draw[ultra thick, draw=olive] (axis cs: 8,1) -- node[above,text=olive]{\footnotesize DNF} (axis cs: 10,1);
      \draw[ultra thick, draw=blue] (axis cs: 8,.2) -- node[above,text=blue]{\footnotesize DNF} (axis cs: 13,.2);
      \draw[ultra thick, draw=red] (axis cs: 8,3) -- node[above,text=red]{\footnotesize DNF} (axis cs: 11,3);
      \addplot+[dashdotted,olive,mark=diamond,mark options={solid,fill=olive},discard if not={rate_per_worker}{\expwordcountrateperworkertwo}] table [x expr=log2(\thisrow{quantum}),y expr=\thisrow{0.999} / 1000000] {\expwordcountquantumwatermarks};
      \addplot+[dashed,blue,mark=o,mark options={solid,fill=blue},discard if not={rate_per_worker}{\expwordcountrateperworkertwo}] table [x expr=log2(\thisrow{quantum}),y expr=\thisrow{0.999} / 1000000] {\expwordcountquantumnotificator};
      \addplot+[solid,red,mark=square,mark options={solid,fill=red},discard if not={rate_per_worker}{\expwordcountrateperworkertwo}] table [x expr=log2(\thisrow{quantum}),y expr=\thisrow{0.999} / 1000000] {\expwordcountquantumtokens};

      \nextgroupplot[
        ymode=log,
        xlabel style={align=center, text width=.30\linewidth},
        xlabel={quantum ($2^x$ns)\\64M tuples/sec},
        ymin=0.1,
        ymax=2,
      ]
      \draw[ultra thick, draw=olive] (axis cs: 8,.4) -- node[above,text=olive]{\footnotesize DNF} (axis cs: 10,.4);
      \draw[ultra thick, draw=blue] (axis cs: 8,.2) -- node[above,text=blue]{\footnotesize DNF} (axis cs: 13,.2);
      \draw[ultra thick, draw=red] (axis cs: 8,1) -- node[above,text=red]{\footnotesize DNF} (axis cs: 11,1);
      \addplot+[dashdotted,olive,mark=diamond,mark options={solid,fill=olive},discard if not={rate_per_worker}{\expwordcountrateperworkertwo}] table [x expr=log2(\thisrow{quantum}),y expr=\thisrow{0.5} / 1000000] {\expwordcountquantumwatermarks};
      \addplot+[dashed,blue,mark=o,mark options={solid,fill=blue},discard if not={rate_per_worker}{\expwordcountrateperworkertwo}] table [x expr=log2(\thisrow{quantum}),y expr=\thisrow{0.5} / 1000000] {\expwordcountquantumnotificator};
      \addplot+[solid,red,mark=square,mark options={solid,fill=red},discard if not={rate_per_worker}{\expwordcountrateperworkertwo}] table [x expr=log2(\thisrow{quantum}),y expr=\thisrow{0.5} / 1000000] {\expwordcountquantumtokens};
    \end{groupplot}
  \end{tikzpicture}
  \caption{Latency for a single-operator ("word-count") dataflow with Flink-style watermarks, Naiad-style notifications, and with \ttokens. We run the workloads on 8 physical cores at three different offered loads. We report median, p999, and maximum latency as we vary the timestamp quantization. Note the different scales on the y axes of the plots.}
  \label{fig:wordcount-quantum}
\end{figure}
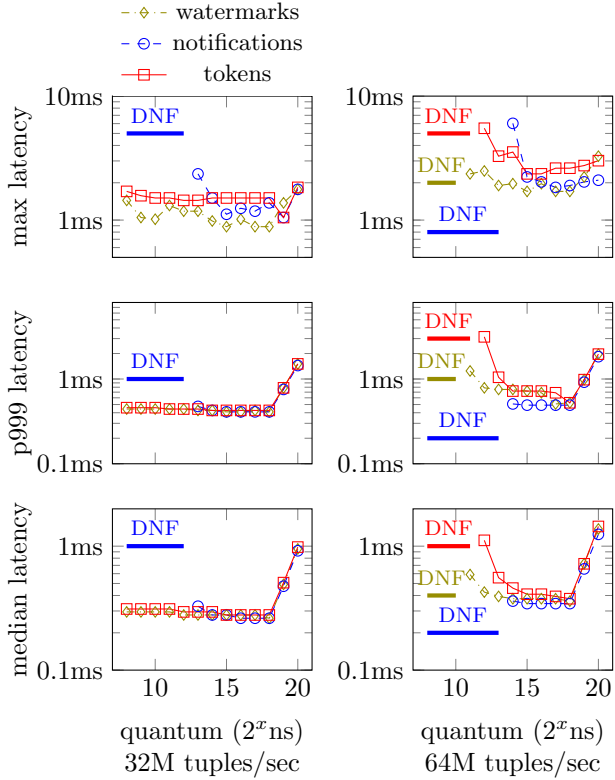

\subsubsection{Varying timestamp granularity}

\Cref{fig:wordcount-quantum} shows the achieved median, p999 (99.9\%), and maximum latency when we vary the granularity of the timestamp quantization under different offered loads: 32 million tuples/sec is below the maximum throughput achievable with fine timestamp granularity by at least some of the coordination mechanisms, and 64 million tuples/sec represent a very high load that all mechanisms cannot sustain with a timestamp quantum of $2^{13}$ns or finer. The performance pattern at lower loads is similar to what we report for 32 million tuples/sec, but with lower latency.

All mechanisms display similar performance characteristics when not overloaded, with two notable exceptions.
First, notifications are unable to handle a timestamp granularity below $2^{13}$ns; this is because they require an interaction between the operator logic and the system for each timestamp. That is not the case for both watermarks and tokens, that can handle any timestamp quantization. Second, the maximum latency for watermarks is 2x smaller than \ttokens for timestamp quantization above $2^{14}$: for this extremely simple single-operator dataflow, watermarks can have slightly lower overhead at the tail.

At very high load (64 million tuples/sec) (i) all mechanisms have significantly higher tail latency and cannot handle the finest timestamp granularities, (ii) both watermarks and \ttokens can handle timestamp granularities finer than notifications, (iii) notifications achieve better p999 (when they are able to sustain the load) possibly due to additional synchronization imposed by the mechanism, and (iv) watermarks display slightly higher median latency at this load.

In this microbenchmark, \ttokens perform essentially on par with watermarks when not overloaded, and behave better when the system is overloaded. Notifications are unable to handle highly granular timestamps in the input data even at lower loads, because every timestamp requires an interaction between the operator logic and the system.


\newcommand{\expwordcountweakscalingworkerrate}   {2000000}
\newcommand{\expwordcountweakscalingwatermarks}   {experiments/wordcount-weak-scaling/epyc7452-6c6aa88-weak-scaling-watermarks.txt}
\newcommand{\expwordcountweakscalingnotificator}  {experiments/wordcount-weak-scaling/epyc7452-6c6aa88-weak-scaling-notificator.txt}
\newcommand{\expwordcountweakscalingtokens}       {experiments/wordcount-weak-scaling/epyc7452-6c6aa88-weak-scaling-tokens.txt}

\newcommand{\expwordcountstrongscalingtotalrate}  {16000000}
\newcommand{\expwordcountstrongscalingwatermarks} {experiments/wordcount-strong-scaling/epyc7452-6c6aa88-watermarks.txt}
\newcommand{\expwordcountstrongscalingnotificator}{experiments/wordcount-strong-scaling/epyc7452-6c6aa88-notificator.txt}
\newcommand{\expwordcountstrongscalingtokens}     {experiments/wordcount-strong-scaling/epyc7452-6c6aa88-tokens.txt}

\begin{figure*}[p] 
  \centering
  \subfloat[Weak scaling. We vary the number of workers and the offered load, which is fixed at 2 million tuples per second per worker.
  Note that Naiad-style notification fail to keep up with load for timestamp quantum = $2^8$ns.] {
  \begin{tikzpicture}
    \begin{groupplot}[
      group style={
        group size=2 by 2,
        horizontal sep=0.2cm,
        vertical sep=0.5cm,
        xticklabels at=edge bottom,
        yticklabels at=edge left,
      },
      xmin=0,
      xmax=17,
      width=.275\linewidth,
      height=.20\linewidth,
      log number format basis/.code 2 args={\pgfmathparse{#1^(#2)}\pgfmathprintnumber{\pgfmathresult}ms},
      max space between ticks=20,
      legend columns=-1,
      legend style={at={(0,1)},anchor=south west,font=\small,draw=none},
      ]

      \nextgroupplot[
        ymode=log,
        ylabel=max latency,
        ymin=0.1,
        ymax=10
      ]
      \addplot+[dashdotted,olive,mark=diamond,mark options={solid,fill=olive},discard if not={quantum}{65536},discard if not={rate_per_worker}{\expwordcountweakscalingworkerrate}] table [x=workers,y expr=\thisrow{max} / 1000000] {\expwordcountweakscalingwatermarks};
      \addplot+[dashed,blue,mark=o,mark options={solid,fill=blue},discard if not={quantum}{65536},discard if not={rate_per_worker}{\expwordcountweakscalingworkerrate}] table [x=workers,y expr=\thisrow{max} / 1000000] {\expwordcountweakscalingnotificator};
      \addplot+[solid,red,mark=square,mark options={solid,fill=red},discard if not={quantum}{65536},discard if not={rate_per_worker}{\expwordcountweakscalingworkerrate}] table [x=workers,y expr=\thisrow{max} / 1000000] {\expwordcountweakscalingtokens};

      \legend{watermarks, notifications, tokens}

      \nextgroupplot[
        ymode=log,
        ymin=0.1,
        ymax=10
      ]
      \draw[ultra thick, draw=blue] (axis cs: 1,.2) -- node[above,text=blue]{\footnotesize notifications DNF} (axis cs: 16,.2);
      \addplot+[dashdotted,olive,mark=diamond,mark options={solid,fill=olive},discard if not={quantum}{256},discard if not={rate_per_worker}{\expwordcountweakscalingworkerrate}] table [x=workers,y expr=\thisrow{max} / 1000000] {\expwordcountweakscalingwatermarks};
      \addplot+[dashed,blue,mark=o,mark options={solid,fill=blue},discard if not={quantum}{256},discard if not={rate_per_worker}{\expwordcountweakscalingworkerrate}] table [x=workers,y expr=\thisrow{max} / 1000000] {\expwordcountweakscalingnotificator};
      \addplot+[solid,red,mark=square,mark options={solid,fill=red},discard if not={quantum}{256},discard if not={rate_per_worker}{\expwordcountweakscalingworkerrate}] table [x=workers,y expr=\thisrow{max} / 1000000] {\expwordcountweakscalingtokens};

      \nextgroupplot[
        ymode=log,
        xlabel style={align=center, text width=.16\linewidth},
        xlabel={workers\\quantum = $2^{16}$ns},
        ylabel=median latency,
        ymin=0.1,
        ymax=1
      ]
      \addplot+[dashdotted,olive,mark=diamond,mark options={solid,fill=olive},discard if not={quantum}{65536},discard if not={rate_per_worker}{\expwordcountweakscalingworkerrate}] table [x=workers,y expr=\thisrow{0.5} / 1000000] {\expwordcountweakscalingwatermarks};
      \addplot+[dashed,blue,mark=o,mark options={solid,fill=blue},discard if not={quantum}{65536},discard if not={rate_per_worker}{\expwordcountweakscalingworkerrate}] table [x=workers,y expr=\thisrow{0.5} / 1000000] {\expwordcountweakscalingnotificator};
      \addplot+[solid,red,mark=square,mark options={solid,fill=red},discard if not={quantum}{65536},discard if not={rate_per_worker}{\expwordcountweakscalingworkerrate}] table [x=workers,y expr=\thisrow{0.5} / 1000000] {\expwordcountweakscalingtokens};

      \nextgroupplot[
        ymode=log,
        xlabel style={align=center, text width=.16\linewidth},
        xlabel={workers\\quantum = $2^{8}$ns},
        ymin=0.1,
        ymax=1
      ]
      \draw[ultra thick, draw=blue] (axis cs: 1,.6) -- node[above,text=blue]{\footnotesize notifications DNF} (axis cs: 16,.6);
      \addplot+[dashdotted,olive,mark=diamond,mark options={solid,fill=olive},discard if not={quantum}{256},discard if not={rate_per_worker}{\expwordcountweakscalingworkerrate}] table [x=workers,y expr=\thisrow{0.5} / 1000000] {\expwordcountweakscalingwatermarks};
      \addplot+[dashed,blue,mark=o,mark options={solid,fill=blue},discard if not={quantum}{256},discard if not={rate_per_worker}{\expwordcountweakscalingworkerrate}] table [x=workers,y expr=\thisrow{0.5} / 1000000] {\expwordcountweakscalingnotificator};
      \addplot+[solid,red,mark=square,mark options={solid,fill=red},discard if not={quantum}{256},discard if not={rate_per_worker}{\expwordcountweakscalingworkerrate}] table [x=workers,y expr=\thisrow{0.5} / 1000000] {\expwordcountweakscalingtokens};

    \end{groupplot}
  \end{tikzpicture}
  }
  \:
  \subfloat[Strong scaling. We vary the number of workers while keeping the offered load fixed at 20 million tuples per second.
  For quantum = $2^{16}$ns, notifications fail to keep up with load with less than 2 workers, and other mechanisms fail with less than 3 workers. For quantum = $2^8$ns, all configurations fail with less than 3 workers, and notifications fail  with any number of workers.] {
  \begin{tikzpicture}
    \begin{groupplot}[
      group style={
        group size=2 by 2,
        horizontal sep=0.2cm,
        vertical sep=0.5cm,
        xticklabels at=edge bottom,
        yticklabels at=edge left
      },
      xmin=0,
      xmax=17,
      width=.275\linewidth,
      height=.20\linewidth,
      log number format basis/.code 2 args={\pgfmathparse{#1^(#2)}\pgfmathprintnumber{\pgfmathresult}ms},
      max space between ticks=20,
      legend columns=-1,
      legend style={at={(-0.2,1)},anchor=south west,font=\small,draw=none},
      ]

      \nextgroupplot[
        ymode=log,
        ylabel=max latency,
        ymin=0.1,
        ymax=10
      ]
      \addplot+[dashdotted,olive,mark=diamond,mark options={solid,fill=olive}] table [x=workers,y expr=\thisrow{max} / 1000000,discard if not={quantum}{65536},discard if not around={total_rate}{\expwordcountstrongscalingtotalrate}] {\expwordcountstrongscalingwatermarks};
      \addplot+[dashed,blue,mark=o,mark options={solid,fill=blue}] table [x=workers,y expr=\thisrow{max} / 1000000,discard if not={quantum}{65536},discard if not around={total_rate}{\expwordcountstrongscalingtotalrate}] {\expwordcountstrongscalingnotificator};
      \addplot+[solid,red,mark=square,mark options={solid,fill=red}] table [x=workers,y expr=\thisrow{max} / 1000000,discard if not={quantum}{65536},discard if not around={total_rate}{\expwordcountstrongscalingtotalrate}] {\expwordcountstrongscalingtokens};

      \legend{watermarks, notifications, tokens}

      \nextgroupplot[
        ymode=log,
        ymin=0.1,
        ymax=10
      ]
      \draw[ultra thick, draw=blue] (axis cs: 1,.2) -- node[above,text=blue]{\footnotesize notifications DNF} (axis cs: 16,.2);
      \addplot+[dashdotted,olive,mark=diamond,mark options={solid,fill=olive}] table [x=workers,y expr=\thisrow{max} / 1000000,discard if not={quantum}{256},discard if not around={total_rate}{\expwordcountstrongscalingtotalrate}] {\expwordcountstrongscalingwatermarks};
      \addplot+[dashed,blue,mark=o,mark options={solid,fill=blue}] table [x=workers,y expr=\thisrow{max} / 1000000,discard if not={quantum}{256},discard if not around={total_rate}{\expwordcountstrongscalingtotalrate}] {\expwordcountstrongscalingnotificator};
      \addplot+[solid,red,mark=square,mark options={solid,fill=red}] table [x=workers,y expr=\thisrow{max} / 1000000,discard if not={quantum}{256},discard if not around={total_rate}{\expwordcountstrongscalingtotalrate}] {\expwordcountstrongscalingtokens};

      \nextgroupplot[
        ymode=log,
        xlabel style={align=center, text width=.16\linewidth},
        xlabel={workers\\quantum = $2^{16}$ns},
        ylabel=median latency,
        ymin=0.1,
        ymax=1
      ]
      \addplot+[dashdotted,olive,mark=diamond,mark options={solid,fill=olive}] table [x=workers,y expr=\thisrow{0.5} / 1000000,discard if not={quantum}{65536},discard if not around={total_rate}{\expwordcountstrongscalingtotalrate}] {\expwordcountstrongscalingwatermarks};
      \addplot+[dashed,blue,mark=o,mark options={solid,fill=blue}] table [x=workers,y expr=\thisrow{0.5} / 1000000,discard if not={quantum}{65536},discard if not around={total_rate}{\expwordcountstrongscalingtotalrate}] {\expwordcountstrongscalingnotificator};
      \addplot+[solid,red,mark=square,mark options={solid,fill=red}] table [x=workers,y expr=\thisrow{0.5} / 1000000,discard if not={quantum}{65536},discard if not around={total_rate}{\expwordcountstrongscalingtotalrate}] {\expwordcountstrongscalingtokens};

      \nextgroupplot[
        ymode=log,
        xlabel style={align=center, text width=.16\linewidth},
        xlabel={workers\\quantum = $2^8$ns},
        ymin=0.1,
        ymax=1
      ]
      \draw[ultra thick, draw=blue] (axis cs: 1,.6) -- node[above,text=blue]{\footnotesize notifications DNF} (axis cs: 16,.6);
      \addplot+[dashdotted,olive,mark=diamond,mark options={solid,fill=olive}] table [x=workers,y expr=\thisrow{0.5} / 1000000,discard if not={quantum}{256},discard if not around={total_rate}{\expwordcountstrongscalingtotalrate}] {\expwordcountstrongscalingwatermarks};
      \addplot+[dashed,blue,mark=o,mark options={solid,fill=blue}] table [x=workers,y expr=\thisrow{0.5} / 1000000,discard if not={quantum}{256},discard if not around={total_rate}{\expwordcountstrongscalingtotalrate}] {\expwordcountstrongscalingnotificator};
      \addplot+[solid,red,mark=square,mark options={solid,fill=red}] table [x=workers,y expr=\thisrow{0.5} / 1000000,discard if not={quantum}{256},discard if not around={total_rate}{\expwordcountstrongscalingtotalrate}] {\expwordcountstrongscalingtokens};

    \end{groupplot}
  \end{tikzpicture}
  }
  \caption{Weak and strong scaling for the word-count workload. Note the different scales on the y axes of the plots. For both weak and strong scaling we report results with the timestamp quantization set to either $2^{16}$ns or $2^{8}$ns.}
  \label{fig:wordcount-scaling}
\end{figure*}
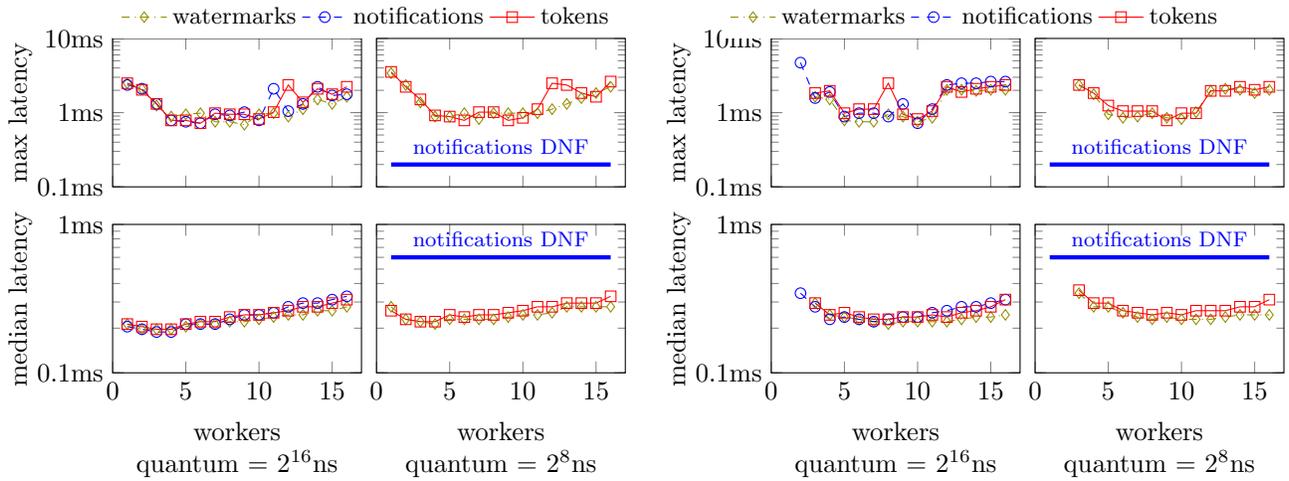

\subsubsection{Scaling}

\Cref{fig:wordcount-scaling} shows the scaling behaviour of the microbenchmark word-count dataflow. At the coarser timestamp quantization granularity, all techniques display nearly identical scaling characteristics. In both strong and weak scaling we can see the system's and techniques' minor inefficiencies starting to affect the reported latency above around 6 workers. At the finer timestamp granularity, Naiad-style notifications fail to keep up with load at any scale, while watermarks and \ttokens display similar behaviour. This demonstrates that \ttokens do not negatively affect scaling.

\newcommand{\expopsequenceseqlengthwatermarksexchange} {experiments/opsequence/epyc7452-6c6aa88-B-vary-sequence-watermarks-exchange.txt}
\newcommand{\expopsequenceseqlengthwatermarkspipeline} {experiments/opsequence/epyc7452-6c6aa88-B-vary-sequence-watermarks-pipeline.txt}
\newcommand{\expopsequenceseqlengthnotificator}        {experiments/opsequence/epyc7452-6c6aa88-B-vary-sequence-notificator.txt}
\newcommand{\expopsequenceseqlengthtokens}             {experiments/opsequence/epyc7452-6c6aa88-B-vary-sequence-tokens.txt}

\newcommand{\expopsequenceweakscalingwatermarksexchange} {experiments/opsequence/epyc7452-6c6aa88-B-scaling-watermarks-exchange.txt}
\newcommand{\expopsequenceweakscalingwatermarkspipeline} {experiments/opsequence/epyc7452-6c6aa88-B-scaling-watermarks-pipeline.txt}
\newcommand{\expopsequenceweakscalingnotificator}        {experiments/opsequence/epyc7452-6c6aa88-B-scaling-notificator.txt}
\newcommand{\expopsequenceweakscalingtokens}             {experiments/opsequence/epyc7452-6c6aa88-B-scaling-tokens.txt}

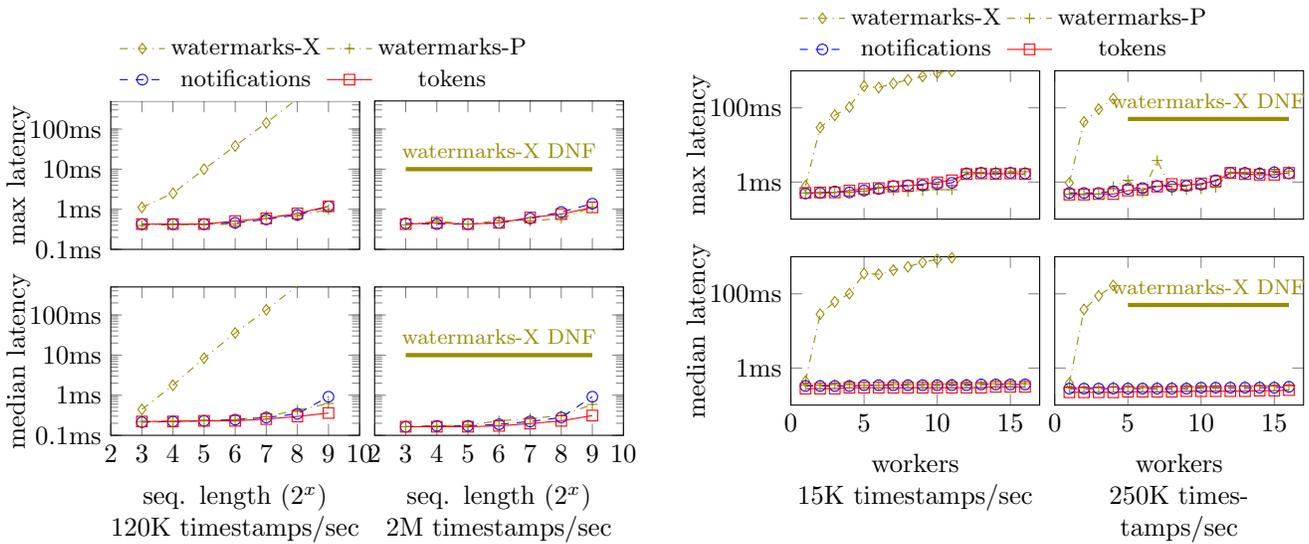
\begin{figure*}[p] 
  \centering
  \subfloat[Sequence of no-op operators. We vary the number of operators in the sequence and the offered load in terms of timestamps/sec. We run the workloads on 8 physical cores.\label{fig:opsequence-length}] {
  \begin{tikzpicture}
    \begin{groupplot}[
      group style={
        group size=2 by 2,
        horizontal sep=0.2cm,
        vertical sep=0.5cm,
        xticklabels at=edge bottom,
        yticklabels at=edge left,
      },
      xmin=2,
      xmax=10,
      width=.275\linewidth,
      height=.20\linewidth,
      log number format basis/.code 2 args={\pgfmathparse{#1^(#2)}\pgfmathprintnumber{\pgfmathresult}ms},
      max space between ticks=10,
      legend columns=2,
      legend style={at={(0,1)},anchor=south west,font=\small,draw=none},
      ]

      \nextgroupplot[
        ymode=log,
        ylabel=max latency,
        ymin=0.1,
        ymax=500
      ]
      \addplot+[dashdotted,olive,mark=diamond,mark options={solid,fill=olive},discard if not={quantum}{65536}] table [x expr=log2(\thisrow{sequence_length}),y expr=\thisrow{max} / 1000000] {\expopsequenceseqlengthwatermarksexchange};
      \addplot+[dashdotted,olive,mark=+,mark options={solid,fill=olive},discard if not={quantum}{65536}] table [x expr=log2(\thisrow{sequence_length}),y expr=\thisrow{max} / 1000000] {\expopsequenceseqlengthwatermarkspipeline};
      \addplot+[dashed,blue,mark=o,mark options={solid,fill=blue},discard if not={quantum}{65536}] table [x expr=log2(\thisrow{sequence_length}),y expr=\thisrow{max} / 1000000] {\expopsequenceseqlengthnotificator};
      \addplot+[solid,red,mark=square,mark options={solid,fill=red},discard if not={quantum}{65536}] table [x expr=log2(\thisrow{sequence_length}),y expr=\thisrow{max} / 1000000] {\expopsequenceseqlengthtokens};

      \legend{watermarks-X, watermarks-P, notifications, tokens}

      \nextgroupplot[
        ymode=log,
        ymin=0.1,
        ymax=500
      ]
      \draw[ultra thick, draw=olive] (axis cs: 3,10) -- node[above,text=olive]{\footnotesize watermarks-X DNF} (axis cs: 9,10);
      \addplot+[dashdotted,olive,mark=diamond,mark options={solid,fill=olive},discard if not={quantum}{4096}] table [x expr=log2(\thisrow{sequence_length}),y expr=\thisrow{max} / 1000000] {\expopsequenceseqlengthwatermarksexchange};
      \addplot+[dashdotted,olive,mark=+,mark options={solid,fill=olive},discard if not={quantum}{4096}] table [x expr=log2(\thisrow{sequence_length}),y expr=\thisrow{max} / 1000000] {\expopsequenceseqlengthwatermarkspipeline};
      \addplot+[dashed,blue,mark=o,mark options={solid,fill=blue},discard if not={quantum}{4096}] table [x expr=log2(\thisrow{sequence_length}),y expr=\thisrow{max} / 1000000] {\expopsequenceseqlengthnotificator};
      \addplot+[solid,red,mark=square,mark options={solid,fill=red},discard if not={quantum}{4096}] table [x expr=log2(\thisrow{sequence_length}),y expr=\thisrow{max} / 1000000] {\expopsequenceseqlengthtokens};

      \nextgroupplot[
        ymode=log,
        xlabel style={align=center, text width=.20\linewidth},
        xlabel={seq. length ($2^x$)\\120K timestamps/sec},
        ylabel=median latency,
        ymin=0.1,
        ymax=500
      ]
      \addplot+[dashdotted,olive,mark=diamond,mark options={solid,fill=olive},discard if not={quantum}{65536}] table [x expr=log2(\thisrow{sequence_length}),y expr=\thisrow{0.5} / 1000000] {\expopsequenceseqlengthwatermarksexchange};
      \addplot+[dashdotted,olive,mark=+,mark options={solid,fill=olive},discard if not={quantum}{65536}] table [x expr=log2(\thisrow{sequence_length}),y expr=\thisrow{0.5} / 1000000] {\expopsequenceseqlengthwatermarkspipeline};
      \addplot+[dashed,blue,mark=o,mark options={solid,fill=blue},discard if not={quantum}{65536}] table [x expr=log2(\thisrow{sequence_length}),y expr=\thisrow{0.5} / 1000000] {\expopsequenceseqlengthnotificator};
      \addplot+[solid,red,mark=square,mark options={solid,fill=red},discard if not={quantum}{65536}] table [x expr=log2(\thisrow{sequence_length}),y expr=\thisrow{0.5} / 1000000] {\expopsequenceseqlengthtokens};

      \nextgroupplot[
        ymode=log,
        xlabel style={align=center, text width=.20\linewidth},
        xlabel={seq. length ($2^x$)\\2M timestamps/sec},
        ymin=0.1,
        ymax=500
      ]
      \draw[ultra thick, draw=olive] (axis cs: 3,10) -- node[above,text=olive]{\footnotesize watermarks-X DNF} (axis cs: 9,10);
      \addplot+[dashdotted,olive,mark=diamond,mark options={solid,fill=olive},discard if not={quantum}{4096}] table [x expr=log2(\thisrow{sequence_length}),y expr=\thisrow{0.5} / 1000000] {\expopsequenceseqlengthwatermarksexchange};
      \addplot+[dashdotted,olive,mark=+,mark options={solid,fill=olive},discard if not={quantum}{4096}] table [x expr=log2(\thisrow{sequence_length}),y expr=\thisrow{0.5} / 1000000] {\expopsequenceseqlengthwatermarkspipeline};
      \addplot+[dashed,blue,mark=o,mark options={solid,fill=blue},discard if not={quantum}{4096}] table [x expr=log2(\thisrow{sequence_length}),y expr=\thisrow{0.5} / 1000000] {\expopsequenceseqlengthnotificator};
      \addplot+[solid,red,mark=square,mark options={solid,fill=red},discard if not={quantum}{4096}] table [x expr=log2(\thisrow{sequence_length}),y expr=\thisrow{0.5} / 1000000] {\expopsequenceseqlengthtokens};

    \end{groupplot}
  \end{tikzpicture}
  }
  \:
  \subfloat[Weak scaling for an operator sequence of 256 no-op operators. We vary the number of workers while keeping the offered load fixed at 15K and 250K timestamps per second, per worker.\label{fig:opsequence-scale}]{
  \begin{tikzpicture}
    \begin{groupplot}[
      group style={
        group size=2 by 2,
        horizontal sep=0.2cm,
        vertical sep=0.5cm,
        xticklabels at=edge bottom,
        yticklabels at=edge left,
      },
      xmin=0,
      xmax=17,
      width=.275\linewidth,
      height=.20\linewidth,
      log number format basis/.code 2 args={\pgfmathparse{#1^(#2)}\pgfmathprintnumber{\pgfmathresult}ms},
      max space between ticks=20,
      legend columns=2,
      legend style={at={(0,1)},anchor=south west,font=\small,draw=none},
      ]

      \nextgroupplot[
        ymode=log,
        ylabel=max latency,
        ymin=0.1,
        ymax=1000
      ]
      \addplot+[dashdotted,olive,mark=diamond,mark options={solid,fill=olive},discard if not={rate_per_worker}{15258}] table [x=workers,y expr=\thisrow{max} / 1000000] {\expopsequenceweakscalingwatermarksexchange};
      \addplot+[dashdotted,olive,mark=+,mark options={solid,fill=olive},discard if not={rate_per_worker}{15258}] table [x=workers,y expr=\thisrow{max} / 1000000] {\expopsequenceweakscalingwatermarkspipeline};
      \addplot+[dashed,blue,mark=o,mark options={solid,fill=blue},discard if not={rate_per_worker}{15258}] table [x=workers,y expr=\thisrow{max} / 1000000] {\expopsequenceweakscalingnotificator};
      \addplot+[solid,red,mark=square,mark options={solid,fill=red},discard if not={rate_per_worker}{15258}] table [x=workers,y expr=\thisrow{max} / 1000000] {\expopsequenceweakscalingtokens};

      \legend{watermarks-X, watermarks-P, notifications, tokens}

      \nextgroupplot[
        ymode=log,
        ymin=0.1,
        ymax=1000
      ]
      \draw[ultra thick, draw=olive] (axis cs: 5,50) -- node[above,text=olive]{\footnotesize watermarks-X DNF} (axis cs: 16,50);
      \addplot+[dashdotted,olive,mark=diamond,mark options={solid,fill=olive},discard if not={rate_per_worker}{244140}] table [x=workers,y expr=\thisrow{max} / 1000000] {\expopsequenceweakscalingwatermarksexchange};
      \addplot+[dashdotted,olive,mark=+,mark options={solid,fill=olive},discard if not={rate_per_worker}{244140}] table [x=workers,y expr=\thisrow{max} / 1000000] {\expopsequenceweakscalingwatermarkspipeline};
      \addplot+[dashed,blue,mark=o,mark options={solid,fill=blue},discard if not={rate_per_worker}{244140}] table [x=workers,y expr=\thisrow{max} / 1000000] {\expopsequenceweakscalingnotificator};
      \addplot+[solid,red,mark=square,mark options={solid,fill=red},discard if not={rate_per_worker}{244140}] table [x=workers,y expr=\thisrow{max} / 1000000] {\expopsequenceweakscalingtokens};

      \nextgroupplot[
        ymode=log,
        xlabel style={align=center, text width=.18\linewidth},
        xlabel={workers\\15K timestamps/sec},
        ylabel=median latency,
        ymin=0.1,
        ymax=1000
      ]
      \addplot+[dashdotted,olive,mark=diamond,mark options={solid,fill=olive},discard if not={rate_per_worker}{15258}] table [x=workers,y expr=\thisrow{0.5} / 1000000] {\expopsequenceweakscalingwatermarksexchange};
      \addplot+[dashdotted,olive,mark=+,mark options={solid,fill=olive},discard if not={rate_per_worker}{15258}] table [x=workers,y expr=\thisrow{0.5} / 1000000] {\expopsequenceweakscalingwatermarkspipeline};
      \addplot+[dashed,blue,mark=o,mark options={solid,fill=blue},discard if not={rate_per_worker}{15258}] table [x=workers,y expr=\thisrow{0.5} / 1000000] {\expopsequenceweakscalingnotificator};
      \addplot+[solid,red,mark=square,mark options={solid,fill=red},discard if not={rate_per_worker}{15258}] table [x=workers,y expr=\thisrow{0.5} / 1000000] {\expopsequenceweakscalingtokens};

      \nextgroupplot[
        ymode=log,
        xlabel style={align=center, text width=.18\linewidth},
        xlabel={workers\\250K timestamps/sec},
        ymin=0.1,
        ymax=1000
      ]
      \draw[ultra thick, draw=olive] (axis cs: 5,50) -- node[above,text=olive]{\footnotesize watermarks-X DNF} (axis cs: 16,50);
      \addplot+[dashdotted,olive,mark=diamond,mark options={solid,fill=olive},discard if not={rate_per_worker}{244140}] table [x=workers,y expr=\thisrow{0.5} / 1000000] {\expopsequenceweakscalingwatermarksexchange};
      \addplot+[dashdotted,olive,mark=+,mark options={solid,fill=olive},discard if not={rate_per_worker}{244140}] table [x=workers,y expr=\thisrow{0.5} / 1000000] {\expopsequenceweakscalingwatermarkspipeline};
      \addplot+[dashed,blue,mark=o,mark options={solid,fill=blue},discard if not={rate_per_worker}{244140}] table [x=workers,y expr=\thisrow{0.5} / 1000000] {\expopsequenceweakscalingnotificator};
      \addplot+[solid,red,mark=square,mark options={solid,fill=red},discard if not={rate_per_worker}{244140}] table [x=workers,y expr=\thisrow{0.5} / 1000000] {\expopsequenceweakscalingtokens};

    \end{groupplot}
  \end{tikzpicture}
  }
  \caption{Impact of a long sequence of operators in the dataflow graph. For Flink-style watermarks we consider two dataflows: one with all-worker exchanges at every stage (watermarks-X) and one where operators form pipelines that are connected locally on each worker (watermarks-P). Note the different scales on the y axes of the plots.}
  \label{fig:opsequence}
\end{figure*}

\subsection{Complex dataflow fragments}
\label{sec:eval-opsequence}

As discussed in \Cref{sec:benefits}, \ttokens do not require continual interaction between the operator and the system to retire timestamps, in particular when an operator is idle for a period of time. To measure the performance benefit of not having to invoke each operator for each successive timestamp, even if no work needs to be performed, we construct a dataflow with a variable sequence of no-op operators (from 8 to 256 no-op operators connected as a sequential pipeline).

\Ttokens and Naiad-style notifications always calculate operator input frontiers (low watermarks) as if each channel between two consecutive operators may exchange data between workers. For Flink-style watermarks we need to distinguish between a scenario where a cross-worker exchange happens at each step (and watermarks are broadcast) and an additional (unrealistic) scenario at the other end of the spectrum where no cross-worker data exchange takes place. A real-world dataflow is likely to have a mix of worker-local and cross-worker channels, and would likely sit somewhere between these two extremes.

\Cref{fig:opsequence} shows the performance impact of handling timestamps for a sequence of idle operators of varying length. \Ttokens, and Naiad-style notifications, and the Flink-style watermark configuration without cross-worker exchange (\textit{watermarks-P}) have almost identical performance that is only marginally affected by the length of the operator chain (\Cref{fig:opsequence-length}) and by the workload scale (\Cref{fig:opsequence-scale}). In this scenario \textit{watermarks-P} has an unrealistic advantage because no coordination information is ever exchanged between workers: each processor operates as a separate unit, and thus does not incur any coordination cost.

When configured to perform exchanges for every inter-operator channel (\textit{watermarks-X}) the latency for Flink-style watermarks degrades linearly with the number of operators in the sequence (\Cref{fig:opsequence-length}) because each operator has to be invoked to forward the watermark which then needs to be broadcast to all other operators. This also fundamentally limits scalability: \textit{watermarks-X} has to process watermarks proportional to the length of the sequence times the number of workers, resulting in high latency even at moderate scale.

By not requiring interaction with each operator for each timestamp, \ttokens matches or outperforms other techniques when handling complex inactive dataflow fragments.

\begin{figure*}[htb]
  \centering

\begin{tabular}{c | c | l l l | l l l | l l l}
  \multicolumn{2}{l}{NEXmark Q4} & \multicolumn{9}{|l}{latency (milliseconds)} \\
  \hline
  & & \multicolumn{3}{| l}{tokens} & \multicolumn{3}{| l}{notifications} & \multicolumn{3}{| l}{watermarks} \\
  tuples/sec & workers & p50 & p999 & max & p50 & p999 & max & p50 & p999 & max \\
  \hline
  4M & 4 & 0.62 & 1.25 & 1.9    & \multicolumn{3}{|c|}{DNF} & 0.25 & 0.59 & 1.25 \\
  4M & 8 & 0.52 & 0.98 & 1.51   & \multicolumn{3}{|c|}{DNF} & 0.29 & 0.56 & 1.44 \\
  4M & 12 & 0.59 & 1.02 & 5.77  & \multicolumn{3}{|c|}{DNF} & 0.38 & 0.56 & 2.49 \\ \hline
  6M & 4 & \multicolumn{3}{|c}{DNF} & \multicolumn{3}{|c}{DNF} & \multicolumn{3}{|c}{DNF} \\
  6M & 8 & 1.31 & 2.62 & 4.19   & \multicolumn{3}{|c|}{DNF} & 0.72 & 2.36 & 4.19 \\
  6M & 12 & 1.25 & 2.36 & 2.88  & \multicolumn{3}{|c|}{DNF} & 0.51 & 1.02 & 3.54 \\ \hline
  8M & 4 & \multicolumn{3}{|c}{DNF} & \multicolumn{3}{|c}{DNF} & \multicolumn{3}{|c}{DNF} \\
  8M & 8 & \multicolumn{3}{|c}{DNF} & \multicolumn{3}{|c}{DNF} & \multicolumn{3}{|c}{DNF} \\
  8M & 12 & 2.03 & 3.93 & 11.53 & \multicolumn{3}{|c|}{DNF} & 0.95 & 2.62 & 3.67 \\ \hline
\end{tabular}

\vspace{.5em}

\begin{tabular}{c | c | l l l | l l l | l l l}
  \multicolumn{2}{l}{NEXmark Q7} & \multicolumn{9}{|l}{latency (milliseconds)} \\
  \hline
  & & \multicolumn{3}{| l}{tokens} & \multicolumn{3}{| l}{notifications} & \multicolumn{3}{| l}{watermarks} \\
  tuples/sec & workers & p50 & p999 & max & p50 & p999 & max & p50 & p999 & max \\
  \hline
  4M & 4 & 0.06 & 0.09 & 0.31 & 0.06 & 0.09 & 0.22 & 0.07 & 0.11 & 0.36 \\
  4M & 8 & 0.06 & 0.1 & 0.46 & 0.06 & 0.09 & 0.41 & 0.08 & 0.13 & 0.66 \\
  4M & 12 & 0.06 & 0.11 & 0.82 & 0.06 & 0.1 & 0.72 & 0.1 & 0.17 & 0.79 \\ \hline
  6M & 4 & 0.06 & 0.1 & 0.23 & 0.06 & 0.1 & 0.38 & 0.07 & 0.11 & 0.26 \\
  6M & 8 & 0.06 & 0.1 & 0.46 & 0.06 & 0.1 & 0.44 & 0.09 & 0.13 & 0.66 \\
  6M & 12 & 0.07 & 0.11 & 0.92 & 0.06 & 0.11 & 0.95 & 0.11 & 0.18 & 0.82 \\ \hline
  8M & 4 & 0.07 & 0.1 & 0.39 & 0.07 & 0.11 & 0.24 & 0.07 & 0.11 & 0.62 \\
  8M & 8 & 0.07 & 0.11 & 0.56 & 0.06 & 0.1 & 0.44 & 0.09 & 0.15 & 0.69 \\
  8M & 12 & 0.07 & 0.11 & 1.02 & 0.07 & 0.11 & 0.92 & 0.11 & 0.19 & 1.31 \\ \hline
\end{tabular}

\caption{End-to-end processing latency for NEXmark query 4 and query 6. We scale the number of workers while keeping the total load fixed at 4, 6, and 8 million tuples/sec. We report median, p999, and maximum latency in milliseconds. For Q4 note that Naiad-style notifications cannot sustain the load for any of the configurations and \ttokens and Flink-style watermarks cannot sustain higher loads with 4-8 workers.}
\end{figure*}

\subsection{NEXMark}
\label{sec:eval-nexmark}

To evaluate \ttokens' performance impact on a realistic, albeit simple, data processing use case, we extended the timely dataflow implementation of the NEXMark queries open sourced\cite{URL:GitHub:Megaphone} by the authors of Megaphone\cite{Hoffmann:2019:MLSMDSD}. The original implementation leverages \ttokens as described in \Cref{sec:stories-scheduling}.
We augmented it by writing the same queries with Naiad-style notifications and Flink-style watermarks.

The NEXMark suite models an auction site in which a high-volume stream of users, auctions, and bids arrive, and standing queries are maintained reflecting a variety of relational queries. For the purporse of this experiment, we focus on queries that result in multi-operator dataflows (Q4 and Q7). Megaphone~\cite{Hoffmann:2019:MLSMDSD} describes the query semantics; for our purposes we only need to highlight that Q4 has a two-stage dataflow where one of the operators handles tokens to calculate a data-dependent windowed maximum, and Q7 has two stateful operators with two consecutive data exchanges. 

\Ttokens avoid the collapse that notifications exhibit for Q4 due to overwhelming numbers of distinct timestamps, and are competitive with watermarks (improving on them slightly for Q7). These queries are relatively simple, only a few dataflow stages, and \ttokens do not have much room to distinguish themselves from watermarks.


\section{Conclusions}

We introduced \ttokens, a coordination primitive for dataflow systems. \Ttokens decouple the sophistication of operator scheduling logic from the task of system-wide coordination.
Operators can add sophistication to their own implementations, including flow control, fine-grained timestamps, and optimized data structures. At the same time, \ttokens simplify the surrounding system, whose role in scheduling no longer needs to be the bottleneck it once was.

Looking forward, we think \ttokens have potential to drive other new dataflow programming idioms, without increasing system complexity.
We are especially interested in \ttokens as dataflow breakpoints, and how holding \ttokens provides external agents the opportunity to suspend execution without fundamentally restructuring dataflow programs.

Finally, we've been delighted by the force multiplier of investing in general dataflow primitives. Many projects quickly and safely implemented new system behavior writing only application-level code. We should have more well-considered primitives and fewer systems.

{
  \bibliography{paper}
}

\end{document}

%% file: shared.tex
\usepackage{float}
\usepackage[font=normalsize]{subfig}
\usepackage{amsthm}
\usepackage{amsmath}
\usepackage{graphicx}
\usepackage{xspace}
\usepackage[utf8]{inputenc}
\usepackage{epsfig,endnotes}
\usepackage{xcolor}
\usepackage{comment}
\usepackage{listings, listings-rust}
\usepackage{tikz}
\usepackage{pgfplots}
\usepackage{ifthen}
\usepackage{smartref}
\usepackage{courier}
\usepackage{chngcntr}
\usetikzlibrary{pgfplots.groupplots}

\usepackage{hyperref}
\hypersetup{
    colorlinks,
    linkcolor={red!50!black},
    citecolor={blue!50!black},
    urlcolor={blue!80!black}
}

\usepackage{cleveref}
\crefname{section}{§}{§§}
\Crefname{section}{§}{§§}

\let\cref\undefined

\usepackage[utf8]{inputenc}
\usepackage[utf8]{luainputenc}

\newcommand{\rom}[1]{(\textit{\lowercase\expandafter{\romannumeral #1\relax})}}

\definecolor{gray}{rgb}{0.5, 0.5, 0.5}
\definecolor{orange}{rgb}{0.88, 0.46, 0.25}
\definecolor{notes}{rgb}{0.4, 0.4, 0.4}
\definecolor{header}{rgb}{0.0, 0.0, 0.8}

\excludecomment{notes}

\newcommand\ttoken{timestamp token\xspace}
\newcommand\Ttoken{Timestamp token\xspace}
\newcommand\ttokens{timestamp tokens\xspace}
\newcommand\Ttokens{Timestamp tokens\xspace}

\newcommand*\circled[1]{\,\tikz[baseline=(char.base)]{
            \node[shape=circle,draw,inner sep=.7pt] (char) {\textcode{#1}};}\,}
\newcounter{clnumcounter}
\counterwithin{clnumcounter}{figure}

\renewcommand{\theclnumcounter}{\Alph{clnumcounter}}
\newcommand{\clnum}{\refstepcounter{clnumcounter}\circled{\theclnumcounter}}
\newcommand{\clref}[1]{\circled{\ref{#1}}}

\pgfplotsset{
        discard if not/.style 2 args={
            filter discard warning=false,
            x filter/.append code={
                \edef\tempa{\thisrow{#1}}
                \edef\tempb{#2}
                \ifx\tempa\tempb
                \else
                    \def\pgfmathresult{NaN}
                \fi
            },
        },
    }

\pgfplotsset{
        discard if not around/.style 2 args={
            filter discard warning=false,
            x filter/.append code={
                \edef\tempcur{\pgfmathresult}
                \edef\tempa{\thisrow{#1}}
                \edef\tempb{#2}
                \pgfmathparse{((\tempb-101)<\tempa)?\tempcur:NaN}
                \edef\tempcur{\pgfmathresult}
                \pgfmathparse{((\tempa-101)<\tempb)?\tempcur:NaN}
            },
        },
    }
    
\newcommand\codesize{\fontsize{8.5pt}{10pt}\selectfont}

\newcommand\textcode[1]{{\codesize\ttfamily{#1}}}